\documentclass{an}
\usepackage{graphicx}
\usepackage{times}
\usepackage{fancyhdr}
\begin{document}
\title{Measuring proton energies and fluxes using EIT (SOHO) CCD areas
outside the solar disk images}
\author{L.V. Didkovsky,\inst{1}\ D.L. Judge, \inst{1}\ A.R. Jones,
\inst{1}\ E.J. Rhodes, Jr.,\inst{1}\ and J.B. Gurman,\inst{2}}
\institute{ University of Southern California, 835 W. 37th Street,
SHS, Los Angeles, California 90089-1341, USA \and NASA Goddard Space
Flight Center, Greenbelt, Maryland 20771, USA }

\date{Received $<$date$>$;
accepted $<$date$>$; published online $<$date$>$}

\abstract {An indirect proton flux measuring tool based on
discrimination of the energy deposited by protons in $128\times128$
pixel EIT CCD areas outside the solar disk images is presented.
Single pixel intensity events are converted into proton incident
energy flux using modeled energy deposition curves for angles of
incidence $\pm$ 60~deg in four EIT spatial areas with different
proton stopping power. The extracted proton flux is corrected for
both the loss of one-pixel events in the range of angles of
incidence as well as for the contribution to the single pixel events
resulting from scattered middle-energy protons (low-energy or
high-energy particles are stopped by the EIT components or pass
through them, accordingly). A simple geometrical approach was found
and applied to correct for a non-unique relation between the
proton-associated CCD output signal and the incident proton energy.
With this geometrical approximation four unique proton incident
energy ranges were determined as 45--49, 145--154, 297--335, and
390--440~MeV. The indirect proton flux measuring tool has been
tested by comparing Solar Energetic Particles (SEP) flux temporal
profiles extracted from the EIT CCD frames and downloaded from the
GOES database for the Bastille Day (BD) of 2000 July 14 and the more
recent 2005 January 20 events. The SEP flux temporal profiles and
proton spectra extracted from the EIT in the relatively narrow
energy ranges between 45 and 440~MeV reported here are consistent
with the related GOES profiles. The four additional EIT extracted
ranges provide higher energy resolution of the SEP data.
 }
\correspondence{leonid@usc.edu}

\maketitle
\section{Introduction}
Major solar flares observed in X-ray and Extreme Ultraviolet (EUV)
radiation allow building and comparing both flare dynamics for the
impulsive phases of flares and their impact on the Earth's
ionosphere, e.g. (Tsurutani et al. 2005). Study of SEP events
produced by those flares, and accelerated by interplanetary shock
formations associated with Coronal Mass Ejections (CMEs), may
clarify the relations between flare (and CME) characteristics, and
the spectra of post-flare events.

If the particle flux spectra have different `fingerprints' for
different flares, then it would be possible to transfer these
differences seen at 1 AU (or, better, seen from a few different
distances to the Sun, (Lin 2005)), and to combine the SEP flux
measurements with corresponding X-ray and EUV flux measurements for
creating a flare's energy spectrum and clarifying the acceleration
mechanism. This task would require both the time-dependent and
distance-dependent particle flux spectra in a number of energy
ranges and an appropriate time-dependent model of particle
acceleration and propagation, e.g. (Tsurutani et al. 2003), as well
as high-cadence measurements in X-ray and EUV bands for the
impulsive phase of the flare. As an initial phase of this kind of
study we would like to start with a number of relatively narrow
energy bands extracted from the SOHO/EIT dark CCD corner areas in
addition to the existing broadband GOES proton flux data.

Particle flux spectra in high-energy ranges may reveal another side
in an analysis of CMEs and SEP (Gopalswamy et al. 2004). The authors
found that ``the active region area has no relation with both the
SEP intensity and CME speed, thus supporting the importance of CME
interaction''. Is there any relation between a flare intensity
(class) and SEP proton spectra?

High-energy particles create a ``noise'' background in a solid state
detector, e.g. (Williams, Arens \& Lanzerotti 1968), strongly
affecting any space-based electronics (Adams, Tsao \& Silberberg
1981). The worst case scenario occurs when the detector consists of
a small number of relatively large pixels. The result of photon
measurements in this case is strongly contaminated by proton-created
electron-hole pairs.

Our goal in this work was to find a way of converting the proton
noise background into a useful signal and to study SEP events in
narrower ranges of energies than available from the NOAA GOES proton
measurements.

SOHO/EIT (Delaboudiniere et al. 1996) uses a $1024\times$1024
thinned, back-illuminated CCD. Scattered protons usually hit
different pixels, depositing either all or a part of their energy in
the CCD active silicon layer, allowing analysis of the density of
the affected pixels and the energy of the detected protons as in the
analysis of a stellar field. To measure the energy of the proton,
some relations between incident proton energy in the range of the
angles of incidence, and the pixel's resulting intensity, are
required. The presence of some optical and mechanical components
through which the detected protons must pass requires that the
stopping power of those components be modeled.

\section{Data observation and the measuring tool}

A few sets of $1024\times$1024 EIT EUV solar images in the 19.5~nm
spectral window were used to extract the proton events. Each set
covers the period of the impulsive phase of the flare and some
pre-flare and post-flare stages. A summary of the data used is given
in Table 1.
\begin{table}[h]
\caption{EIT 19.5~nm data observations.} \label{tlab}
\hspace{0pt}
\begin{tabular}{ccc} \hline
  Date & Images & Cadence \\
   &  & (min)           \\
\hline
2000 Jul 14 & 91 & 15.6  \\
\hline
2000 Jul 15 & 71 & 19.9  \\
\hline
2005 Jan 20 & 93 & 15.2 \\
\hline
\end{tabular}
\end{table}

The information about proton events was extracted from two (N-W and
S-W) corner areas of the CCD images (see subsection 5.1 and Figure
6), each of $128\times$128 pixels in size. These areas are outside
of centered solar images but consist of some bright coronal features
visible in the EUV, and some instrumental sources of noise, like a
bright unfocussed grid associated with the filter support mesh
structure, which develops during an impulsive phase of a solar
flare. To reduce the contamination of the proton-related signals by
these features, a spatial filtering (result = original data minus
filter) was applied to each image of the series after subtracting
dark frames. The size of the filter's window was selected as minimal
as possible, 3 pixels, to effectively filter out all low spatial
frequencies of the unwanted background.

\subsection{One-pixel SEP events}

Clearly, the only correct interpretation of the pixel-based
intensity detector signal is when the relation ``one particle - one
pixel'' is used. In the other two cases when either one particle
hits two (or more) pixels or two (or more) particles deposit energy
in the same pixel, the energy deposited in the CCD pixel is shared
between those events in some unknown proportion, making the planar
CCD an inappropriate detector.

The first of these two cases may be easily detected and eliminated
using an appropriate data reduction algorithm.  The algorithm we
developed is specifically designed to filter out all but
single-pixel events.

The second case may be considered unimportant too in our analysis of
relatively small fluxes, collecting times, and the size of the
pixels. For any analyzed range of SEP energy the rate of one-pixel
events, registered during the EIT collecting time, is not larger
than 0.5 percent, and is substantially smaller for high-energy
ranges. The statistical error introduced by the ``two particles -
one pixel'' events is less than the detecting sensitivity of this
method, about 0.013 particles/cm${^2}$/s/sr/MeV. This means that the
EIT CCD detector can be used to analyze the energies of an incident
particle quasi-isotropic distribution.

The EIT CCD geometry (21~$\mu$m square pixels with the thickness of
the active silicon layer d=12~$\mu$m) (Moses 2004) determines the
opening of both front and back-side cones for incoming particles to
create the one-pixel events. The axis of the cones coincides with
the EIT optical axis and have a maximal opening of $\pm$ 68~deg.
Only about 10${^{-3}}$ $\%$ or less of all quasi-isotropic
particles could 
still hit a pixel in its corner points and produce the one-pixel
event. For a simplified case of this calculation with axial symmetry
and a cylindrical pixel (D = 21~$\mu$m) the maximal opening is about
$\pm$ 60.25~deg. We have limited the outer edge of the opening to
$\pm$ 60~deg with the amount of the one-pixel events at this edge of
about 1.5 $\%$. In contrast with this, about 100 $\%$ of all
particles of the quasi-isotropic flux could produce the one-pixel
events at normal incidence.

Obviously, any particles whose trajectory is inclined with respect
to a normal incidence path has a probability to create two or more
pixel events. This probability for the micro-isotropic fluence
(isotropic over the size of the area of detectors used) is directly
related to the CCD geometry and is a linear function of the angle of
incidence with probability P(0~deg)=0 and P(60~deg)=0.985. It was
used to correct the underestimated one-pixel SEP flux events over
the two-pixel events for any inclined particle path, see $k_{1}$ in
the Equation (2).

\subsection{Four spatial areas of the EIT}

In the range of incident proton trajectories which create one-pixel
events ($\pm$ 60~deg) we have determined four spatial areas with
either different opto-mechanical EIT components and substantially
different stopping power or with a large range of the angles of
incidence. Each of the four areas requires a separate model of the
proton energy deposition relations. These relations may be
represented either by a unique curve with very small deviations on
the ranges of the area or by two distinctive curves corresponding to
the edges of the area and showing a wider range of energy deposition
relations. These areas are (Figure 1): (i) an angular opening $\pm$
2~deg through the secondary mirror and some mechanical components;
(ii) a space between the secondary (SM) and primary mirror (PM) with
opening $\pm$ 2 -- 8~deg through the hole in the PM; (iii) the space
through the PM with opening of $\pm$ 8 -- 18~deg; and (iv) the space
between the outer diameter of the PM and the chosen outer edge of
the one-pixel event openings, $\pm$ 18 -- 60~deg.
\begin{figure}
\resizebox{\hsize}{!} {\includegraphics[width=26pc]{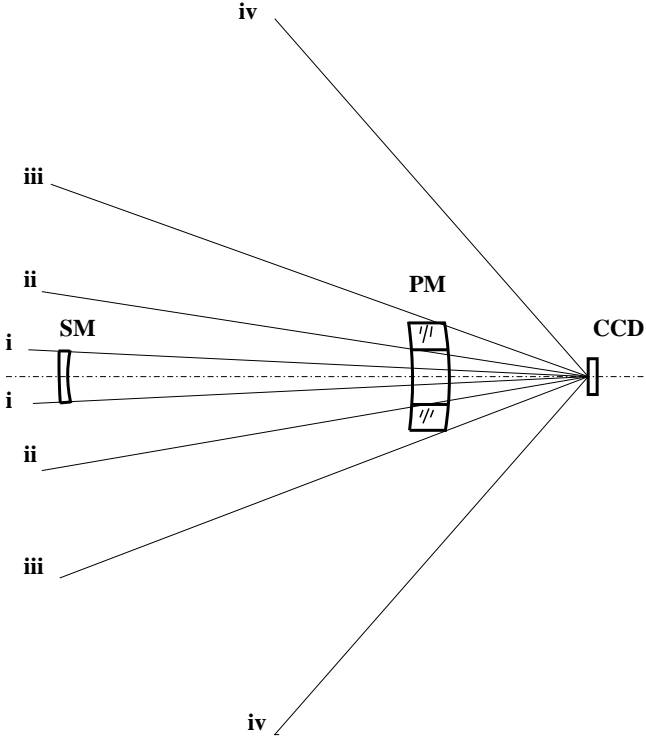}}
\caption{ A simplified schematic view of the EIT (not to scale) and
chosen four spatial area edges i-i, ii-ii, iii-iii, and iv-iv with
different conditions in these areas for deposition of proton energy
to the CCD pixels.}
\end{figure}

Five curves for determining proton deposited energy (PDE) in the CCD
at incidence angles 0, 5, 13, 18, and 60~deg and three curves for
proton stopping power (PSP) at 0, 8, and 18~deg were modeled. The
PSP for the area ii-iii (through the PM) was modeled by two curves
to determine changes of the stopping power associated with the
changes of the angle of incidence between 8 and 18~deg. The
calculated PSP were used to make a corresponding correction to the
proton incident energy when the relations of the energy deposited in
the CCD versus the angle of incidence were finally modeled. For the
three central areas with the opening of $\pm$ 0 -- 18~deg we have
assumed determining PDE for the middle angles of these area angular
ranges only, because changing of the angle of incidence from 0~deg
(first area, center) to 18~deg (third area, outer edge) just
slightly changes the effective length of the mean path through the
CCD silicon layer of 0.6 $\mu$m, which makes corresponding curves
practically overlapped. For the area (iii -- iv) we calculated PDE
for both edges of its opening to have a corresponding range of PDE.

For the same cone openings from the back side of the EIT (the
spacecraft side) we have assumed that the quite thick CCD housing
and spacecraft components represent a substantial stopping power for
the proton flux, which is directed backward, to the Sun. This
stopping power is sufficient to stop protons with lower energies or
substantially shift their incident energy to a lower energy range.
The typical proton spectra in the MeV energy ranges (e.g. GOES data)
show a substantial decrease of the proton flux when the energy is
changed from a lower to a higher level. It means that high-energy
protons that could still come through the back side of the EIT would
add to the normally intense low-energy flux detected from the front
side of the EIT, but no more than a small fraction (1-2 $\%$) or
less of the ``weakened'' high-energy protons. If this assumption is
correct, the SEP flux in the low-energy bands should be slightly
higher than the one measured by dedicated instruments. This error
should be maximal (a few percent) during the short initial time of
the SEP flux event, corresponding to the high-energy proton flux
peak. Moreover, because the arriving time is different
(higher-energy particles arrive earlier than low-energy ones), this
error related to the back-side particles could create a peak in the
low-energy temporal SEP flux profile a while before the pure
low-energy peak. The SEP proton fluxes extracted from the EIT show
(see section 3) that the assumption to ignore the back-side
particles is correct within the error of about 1-2 $\%$.

Modeling of energy deposited by protons in the CCD active silicon
layer was based on the Stopping and Range of Ions in Matter program
(SRIM) (Ziegler \& Biersack 2003). Both PSP and PDE were calculated
with the Monte-Carlo statistics for 5000 protons each. As the SRIM
material library does not contain the zerodur glass ceramic ${\rho}
= 2.53~g/cm^{3}$ used for producing the EIT mirrors, the closest
appropriate optical material - quartz ${\rho} = 2.32~g/cm^{3}$ and
silicon ${\rho} = 2.32~g/cm^{3}$ (for a reference) were used for
modeling the stopping power with practically the same result. The
EIT envelope with thickness of about 75 mils (1.9~mm) (Newmark 2005)
of aluminum was considered with variable thickness as a function of
the angle of incidence. The energy deposited by recoils in the
silicon layer was ignored as it was a few orders of magnitude less
than that deposited directly by ions. Figure 2 shows the modeled
energy deposition curves (converted into DNs) for the four spatial
areas of the EIT (Fig 1), where area 1 is i-i, area 2 is i-ii, area
3 is ii-iii, and area 4 is iii-iv.


 \begin{figure}[h]
 \resizebox{\hsize}{!} {\includegraphics[width=26pc]{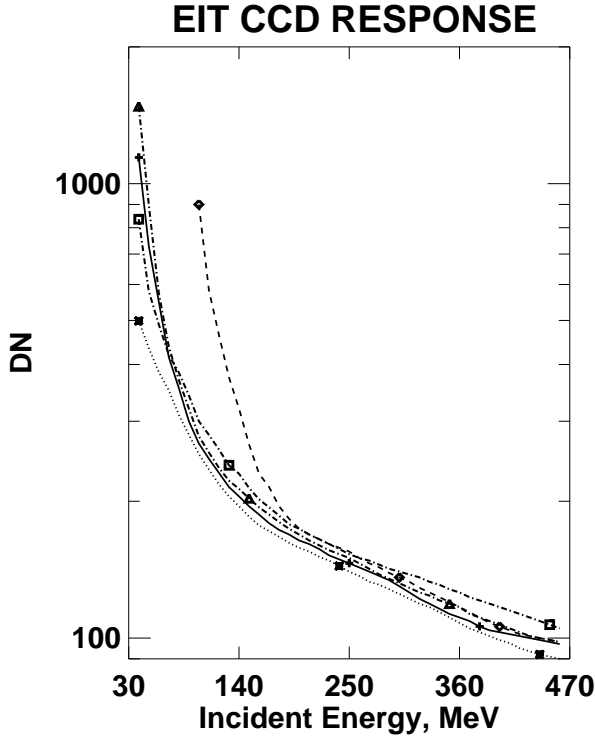}}
 \caption{EIT CCD camera Discrete Numbers (DN) as a function of
proton incident energy for the four EIT opening areas (Fig 1) marked
with pluses (solid line for area 1), asterisks (dotted for area 2),
diamonds (dashed for area 3), triangles and squares (dash-dotted
lines for area 4 at 18 deg and 60 deg, correspondingly).}
   \end{figure}

Energy deposited in the CCD $E_\mathrm{d}$ (not shown in Figure 2)
was converted into the CCD camera output signal (DN) using the
following relation.
\begin{equation}
DN_\mathrm{i} = E_\mathrm{d,i}/E_\mathrm{p}/CCD_\mathrm{s}
\end{equation}

where  $E_\mathrm{p}$ is the deposited energy required to create an
electron-hole pair, $E_\mathrm{p}$ = 3.65 $\mathrm{eV}$;
$CCD_\mathrm{s}$ is the sensitivity of the CCD camera,
$CCD_\mathrm{s}$ = 18~$\mathrm{e}/DN$; $\mathrm{i}$ is the spatial
area index.

The CCD sensitivity $CCD_\mathrm{s}$ (Delaboudiniere et al. 1996)
corresponds to the 18~$\mathrm{e}/DN$ measured at the launch time.
We have assumed that even if the CCD sensitivity has changed due to
solar EUV radiation of the illuminated solar disk pixels, the corner
(dark) area pixels used for this analysis should all have about the
same pre-launched sensitivity.

Figure 2 shows that the whole range of proton incident energies has
two distinctive sub-ranges, namely between 40 and 180~MeV and
between 180 and 460~MeV. The difference between these two sub-ranges
is reflected by the different CCD response of the two sub-ranges. In
the first sub-range both the stopping power in each of the four
spatial areas and different angles of incidence strongly affect the
amount of energy deposited due to the high level of interactions
between the low-energy protons and the active silicon layer. The
influence of these two factors in the second sub-range is relatively
small, because high-energy protons transit the silicon layer with a
lower level of interactions. Figure 2 also shows that the first
sub-range has two distinctive portions, 40--110 and 110--180~MeV.
The lowest energy portion represents protons penetrating through EIT
components in all areas, except the area through the PM. The EIT PM
together with the EIT envelope can stop all protons if their energy
is below 110~MeV and angles of incidence are $\pm$ 8 -- 18~deg.

Because of the wide area of the energy deposition curves, one can
see that the CCD response with DN equal to, e.g., 490 may be created
by protons with energies between 42~MeV (area i--ii) and 120~MeV
(area ii--iii). The real difficulty of any direct approach to
convert the DNs into proton incident energies could come from the
fact that protons with any exact energy in the range of 42--120~MeV,
determined above, might produce a wide range of DNs as a function of
the angle of incidence (the spatial area). For example, protons with
the incident energy of 115~MeV (a vertical line at 115~MeV) would
have the response from 240 to 690 DNs. This means that the direct
method of converting proton incident energy into the CCD output
signal (Figure 2) has no unique solution, thus producing the signal
contamination. To solve this problem a simple geometrical approach
has been found. It is described in the next subsection.

\subsection{Incident SEP energy ranges and corresponding DNs}

Figure 2 shows that the task of extracting the information about the
incident proton energy flux from the CCD intensity signal is not a
trivial one. A vertical line, which corresponds to a given incident
proton energy on Fig 2 intersects a number of energy deposition
curves and shows a corresponding range of DNs, making the task of
interpretation of the results of measurements clearly uncertain.
Nevertheless, in an approximation of a geometrical approach, shown
in Figure 3, this task can be solved.

\begin{figure}[h]
\resizebox{\hsize}{!} {\includegraphics[width=26pc]{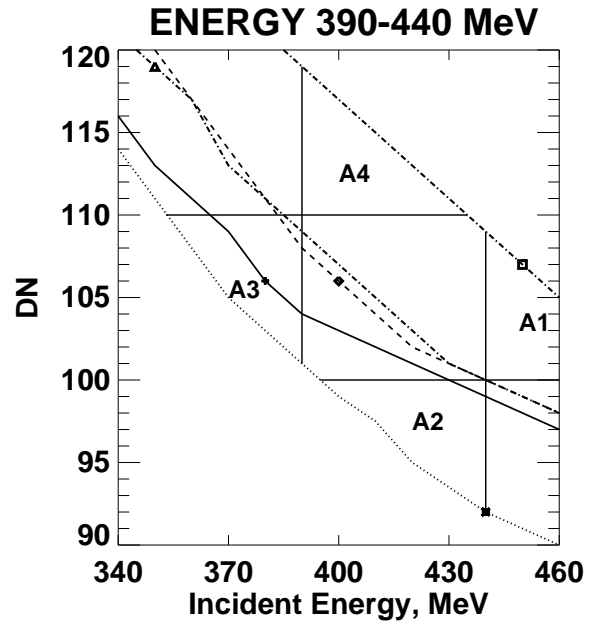}}
   \caption{An enlarged fragment of Fig 2 used for determination of the highest energy
   range (see Table 2). The determined energy range between two vertical lines at 390 and 440 MeV
   leads to a minimal signal contamination if combined with the appropriate
   signal range between two horizontal lines at DN1 = 100 and DN2 = 110.
   The condition to totally avoid
   the signal contamination is to make areas A1 and A2 as well as A3 and A4
   equal to each other. The triangle shaped areas A1--A4 are created by
   two marginal energy deposition curves (marked with asterisk and square)
   and crossed straight lines. }
   \end{figure}

Figure 3 shows a portion of the energy deposition area (high-energy
part of Fig 2) between two marginal energy deposition curves marked
with an asterisk (dotted line for area 2) and a square (dash-dotted
line for area 4 at 60 deg). A combination of two horizontal lines
with two vertical lines shows the condition when two of the four
triangle shaped areas A1 and A3 may be virtually transferred,
replacing other two areas A2 and A4 if the areas A1 and A2, as well
as A3 and A4, are equal to each other. The virtual transferring
works when, e.g., the area A1, which adds to the CCD signal between
DN1 = 100 and DN2 = 110 but should not add in the E1 = 390~MeV
through E2 = 440~MeV energy range, is equal to A2, which does not
add to the CCD signal between DN1 and DN2 but should be counted in
the E1--E2 energy range. With the virtual transfer, when the A1 (A3)
area ``replaces'' the A2 (A4), the appropriate range of DNs between
DN1 and DN2 corresponds to the incident energy range from E1 to E2.

As an example of using the geometrical approach we have determined
four incident proton energy ranges and corresponding four ranges of
DNs (Table 2). The number of determined energy bands may be larger
than the four we have used in this analysis if they are consistent
with the requirement of the geometrical approach found.

\begin{table}[h]
\caption{Deposited energies $E_\mathrm{d}$ and DNs for incident
energy subranges $E_\mathrm{i}$.} \label{tlab}
\begin{tabular}{ccccc}
\hline
$E_{1-2}$ & Median $E_\mathrm{i}$ & $\Delta{E}_\mathrm{i}$ & $E_\mathrm{d}$ & $DN_\mathrm{i}$ \\
 (MeV)  & (MeV)   & (MeV)         & (keV)     & \\
\hline
45--49   & 47   & 4 & 85.4--62.4 & 1300--950  \\
\hline
145--154  & 150 & 9 & 15.2--14.8   & 232--225 \\
\hline
297--335 & 316  & 38 & 8.9--8.2   & 135--125  \\
\hline
390--440 & 415  & 50 & 7.2--6.6   &  110--100  \\
\hline
\end{tabular}
\end{table}

\subsection{High-energy SEP fluxes}

The proton flux temporal profiles for any of the four energy
subranges $E_\mathrm{i}$ were found as:
\begin{equation}
F_\mathrm{i}(E_\mathrm{i},t)=k_\mathrm{1}*N_\mathrm{i}(E_\mathrm{i},t)/S/T/\Delta{E}_\mathrm{i}/\alpha
- k_\mathrm{2}
\end{equation}

\(N_\mathrm{i}(E_\mathrm{i},t)\) is a mean sum of pixels in a given
subrange of Discrete Numbers $DN_\mathrm{i}$:
\begin{equation}
N_\mathrm{i}(E_\mathrm{i},t)=0.5\sum_{DN_{i}}P_\mathrm{1,i}+
P_\mathrm{2,i}
\end{equation}

where $k_\mathrm{1}$ = 2, a statistical coefficient to cover the
loss of the one-pixel events underestimated due to converting them
into two-pixel events at the angles of incidence larger than
0.0~deg; S is the area of $128\times$128 pixels, S = 0.0723
cm${^2}$; T is the mean integration time (12.6 s), which includes
the camera exposure time plus shutter operation time (shutter is
transparent to high energy protons); $\Delta{E}_\mathrm{i}$ is a
number of MeV for a given energy subrange; $\alpha$ = 0.84 sr
corresponds to the cone opening of $\pm$ 30~deg. This opening was
determined as statistically equivalent to the whole opening of $\pm$
60~deg if the probability for both one-pixel and two-pixel events in
this smaller area could be constant and equal to 1.0 (equal to zero
outside of this area), instead of real uniform (linear) distribution
of probabilities from 1.0 to 0.015 and from 0.0 to 0.985 for
one-pixel and two-pixel events at 0 and 60~deg, accordingly;
$DN_\mathrm{i}$ is the range of DNs corresponding to the
contamination-free range of $E_\mathrm{i}$, (Table 2);
$P_\mathrm{1-2}\mathrm{i}$ are corresponding numbers of
proton-pixels events in two S-areas for the energy subrange
$E_\mathrm{i}$; $k_\mathrm{2}$ is a statistical coefficient to
correct the flux over produced by scattered on the EIT mechanical
components particles initially coming outside the one-pixel opening.
This coefficient depends of the intensity of the SEP flux and is 0.4
for the BD and 2005 January 20 events. The approximate relation we
inferred to determine this coefficient for intense SEP flux events
is
\begin{equation} k_\mathrm{2} = \ln{({^4}\sqrt{F_{80-165}})}
\end{equation}
where $F_{80-165} \geq 1.0 $ is GOES maximal SEP event flux in the
energy range of 80--165~MeV.

\section{Temporal EIT and GOES SEP flux profiles}

The EIT-based SEP flux measuring tool was tested with two
geo-effective solar flare events (Table 1) and compared to the
available GOES database temporal profiles taken in different and
lower resolution energy ranges. The GOES energy ranges used for
comparison are 40, 80--165, and 165--500~MeV. The results of the
calculation of proton fluxes for the analyzed flare events are shown
in Figure 4.

Because of the quite small size of the corner CCD areas, each of
$128\times128$ pixels, which is equivalent to 0.072 cm${^2}$ area,
the lower value for SEP fluxes in Figure 4 is limited for both EIT
extracted fluxes as well as the GOES fluxes, to 0.003
particles/cm${^2}$/s/sr/MeV, which is four times lower than the
sensitivity of the EIT proton flux measuring tool (0.013
particles/cm${^2}$/s/sr/MeV). This limit allows some pre-flare
details of the GOES flux to be seen.

GOES data (GOES-8 for 2000/07/14 and GOES-11 for the 2005/01/20
event) were over-plotted on Figure 4 with thin lines for a
reference. Even with different energy band resolution one can see
that both EIT and GOES temporal profiles show a good match in both
the flux maximal values and the temporal profiles. Some differences
in the level of measured SEP flux are consistent with the known
decrease of the flux toward the high-energy ranges. Specifically,
extracted EIT SEP flux, e.g., for the 45--49~MeV energy range (mean
is 47~MeV) is lower than that measured by GOES for 40~MeV SEP. Some
fast changes in the EIT SEP flux temporal profiles are lost due to
the 13--20 min image cadence, compared to the 1 min cadence in the
GOES data.

\begin{figure}[h]
\resizebox{\hsize}{!} {\includegraphics[width=26pc]{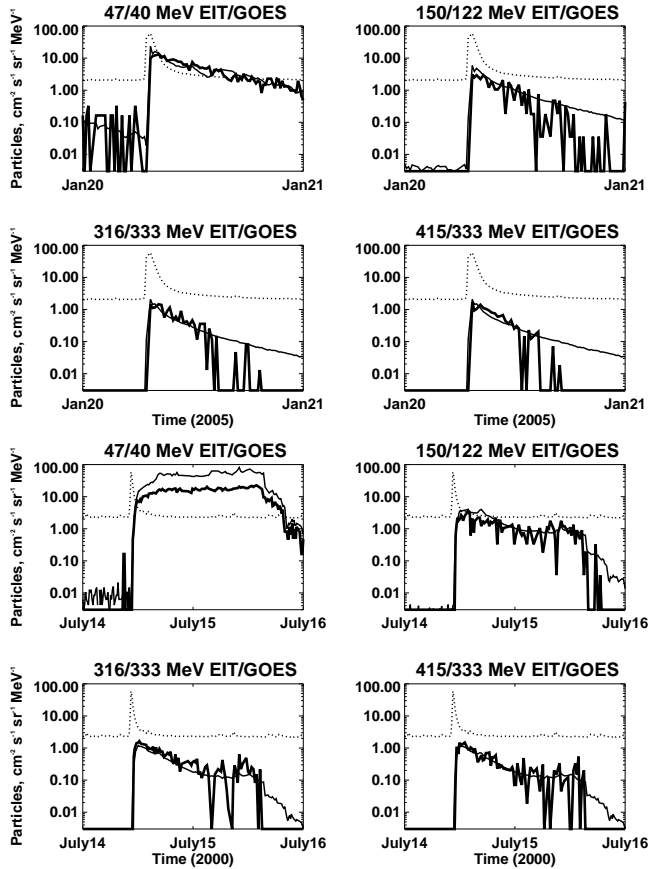}}
   \caption{
Proton fluxes (particles, 1/cm${^2}$/s/sr/MeV) extracted from EIT
(thick line) and from GOES database (thin line) for 2005 January 20
and 2000 July 14 events. The dotted line shows GOES X-ray data in
arbitrarily units with 1-min cadence.}
   \end{figure}

Temporal EIT and GOES flux profiles in the high-energy ranges show
more differences than one can see between the 47 and 40~MeV
profiles. When the number of proton pixel-events registered by the
EIT CCD drops to a few as the high-energy flux decreases, the EIT
temporal profiles show fluctuations to the zero flux level. The
important detail is that when the flux value resumes, it corresponds
to the whole energy range's trend.

Figure 4 has over-plotted GOES X-ray (0.1--0.8~nm) temporal profiles
to show the time delays between the X-ray peak of the solar flare
and corresponding SEP flux peaks. These time delays may be used to
study relativistic features of some post-flare SEP events with a
larger number of energy ranges than is available from the GOES
database.

\section{EIT and GOES proton flux spectra}

An important detail of the current work is to show that even four
additional, but narrower, EIT energy ranges substantially reduce the
uncertainty of the ``flux vs energy'' proton spectra built with
available GOES measurements. Figure 5 shows GOES and EIT proton
spectra.

\begin{figure}[h]\
\resizebox{\hsize}{!} {\includegraphics[width=26pc]{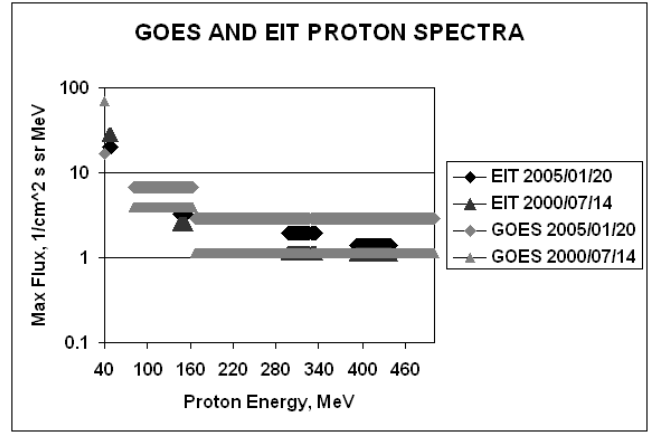}}
  \caption{
GOES and EIT proton spectra for the 2000/07/14 and 2005/01/20 SEP
events. The length of the horizontal bars reflects the energy
range.}
   \end{figure}

Proton spectra for both GOES and EIT data are in good agreement with
each other. The GOES maximal flux values in the energy ranges
165--500~MeV for the analyzed events shown in Figure 5 as long bars
are updated by higher resolution (shorter) EIT bars. The EIT bars
make it possible to determine slopes of the proton spectra in the
energy range of 145--440~MeV, which is not possible with the GOES
long horizontal bars. The statistics for two analyzed events is
quite small when attempting to develop some common slope
characteristics for solar flare events. Nevertheless, an interesting
feature of Figure 5 is that the ``biggest proton event since 1989''
after the X-7 solar flare of 2005 January 20 in the NOAA 10720 AR
shows a substantially smaller total decrement than the one for the
BD event with about the same slope in the high-energy range.

\section{Discussion}

\subsection{Instrumental issues}

Proton flux temporal profiles extracted from EIT and compared to the
GOES-based profiles show good matching of time-variable changes for
lower energy ranges. Higher energy flux profiles are more affected
by some small additional flux due to scattering particles coming
from the outside of the one-pixel opening. The number of scattered
particles is directly related to the intensity of SEP flux event in
the middle range of energies, e.g., GOES 80--165~MeV energy range
(4). Lower and higher energy particles do not affect the
measurements because they are either absorbed or go through the EIT
components. The correction of the SEP flux by subtracting the
``scattered'' portion estimated as $k_{2}$ (4) develops itself
mostly for the low-intense post-flare portions of the SEP flux
temporal profiles not analyzed in this work.

Another feature of the EIT measuring tool is some spatial asymmetry
in the one-pixel opening cone of $\pm$ 60~deg. Four $128\times$128
pixel corner CCD areas show two quite different pairs of the
extracted proton flux signals. The signals extracted from the first
pair, which consists of NW and SW pixel areas are in good agreement
to each other and to the modeled energy deposition relations. The
temporal flux profiles extracted from the second pair of the CCD
pixel areas (NE and SE) are substantially different. They show some
lower incident energy range than that correspondent to the range of
DNs. This shift of the energy range may be caused by some spatial
asymmetry of the EIT regarding its optical axis in the range of the
one-pixel events opening of $\pm$ 60~deg. The most likely the
difference is caused by the stopping power of other SOHO
instruments. We have estimated this asymmetry comparing SEP flux
temporal profiles for each of the four corner areas of the CCD. The
extracted flux profiles from two of the four CCD areas, which
correspond to the modeled energy deposition curves, were used for
our analysis of the SEP events. The flux signal from other two CCD
areas does not correspond to the modeled deposition energy curves
and was excluded from the analysis. Figure 6 shows for 2000 January
20 event flux profiles marked with the thick line for the western
CCD areas (calibrated and used), and with the thin line for the
eastern CCD areas (we did not calibrate and use them).

\begin{figure}[h]\
\resizebox{\hsize}{!} {\includegraphics[width=26pc]{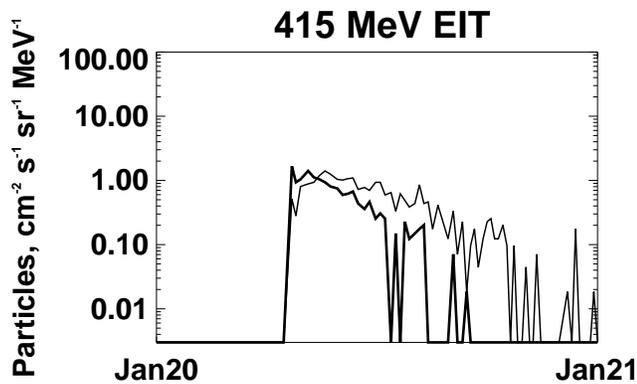}}
\caption{ Proton flux temporal profiles for 2005 January 20 event.
The mean extracted from the EIT flux for the two western corner
areas (thick line) is compared with the flux from the two eastern
areas (thin line) in the energy range of 390--440~MeV. }
   \end{figure}

The thin line (Figure 6) shows slower rise, a delayed maximum,
larger flux, and smaller decrement, all typical features of a lower
energy flux.

\subsection{Temporal SEP flux profiles and proton spectra}

The proton fluxes taken indirectly from EIT images in the four
energy ranges of 45--49, 145--154, 297--335, and 390--440~MeV
(Figure 4) were used to build proton spectra. All these spectra
required extracting proton flux peak-values determined well above
the lower limit of sensitivity of the EIT proton flux measuring
tool, where the temporal profiles match corresponding GOES profiles
and the statistical confidence is high.

Proton spectra showed two distinctive energy ranges clearly seen in
the Figure 5, where fluxes extracted in EIT narrower energy ranges
and compared to the GOES allowed seeing details of the high-energy
spectra. Proton spectra in the high-energy range of 145--440~MeV
show that the slopes for the two analyzed SEP events are quite small
and similar. Lower energy range data for particles between 40 and
145~MeV demonstrates different trends for the analyzed events but
this result should be verified with analysis of a larger number of
SEP events.

If the lower energy proton spectra demonstrate different slopes, it
may reveal either a flare specific or CME specific source of
acceleration.

\section{Conclusions}

A proton flux measuring tool based on using a planar CCD detector
was proposed, realized, and tested by comparing proton flux temporal
profiles and proton spectra extracted from the EIT with those taken
from the GOES database. Temporal profiles (fluxes, shapes, and
decrements) in the energy ranges of 45--49, 145--154, 297--335, and
390--440~MeV extracted from EIT correspond to or match available
profiles from the NOAA GOES database in the energy ranges of 40,
80--165, and 165--500~MeV. Combined EIT-GOES proton spectra show
much more detail when the GOES low energy resolution data are
updated with the corresponding extracted EIT data in narrower energy
ranges.

Proton spectra extracted from EIT for the analyzed SEP events show
two distinctive energy regions, high- and low-energy, with different
spectral features. The slopes of the proton spectra in the
high-energy region (145--440~MeV) are quite small and similar to
each other. In the lower energy range, between 40 and 145~MeV,
slopes are substantially different for each analyzed event. If
confirmed by a larger statistics, similar slopes of the proton flux
spectra for the events with different energy releases may be strong
evidence that the high-energy proton distribution does not depend on
a flare class but follows a similar acceleration law, e.g., produced
by a common post-flare acceleration (propagation) topology related
to CME shock waves. Different slopes for the low-energy region may
be related to either a flare specific or CME specific source of
acceleration.

\begin{acknowledgements}
SOHO is a project of international cooperation between ESA and NASA.
The Authors would like to thank Dan Moses and Frederic Auchere for
some details of the EIT design. We are also grateful to Sasha
Kosovichev and Linton Floyd for their comments and suggestions.
\end{acknowledgements}


\begin{thebibliography}{}

\bibitem{Adams81}
Adams, J.H., Tsao, C.H. and Silberberg, R.:1981, `Cosmic Ray Effects
on Electronics, Part I: The Near-Earth Particle Environment', {\it
NRL Report\/}, {\bf 4506.}

\bibitem{Delaboudiniere96} Delaboudiniere, J.P., Artzner, G.E., Brunaud, J.
et al.: 1996, `EIT: Extreme-Ultraviolet Imaging Telescope for the
SOHO Mission', {\it Solar Physics\/} {\bf Vol.~162}, p.~291.


\bibitem{Gopalswamy04}
Gopalswamy, N., Yashiro, S., Krucker, S., and Howard, R.A.: 2004,
`CME Interaction and the Intensity of Solar Energetic Particle
Events', {\it In Proc. of IAU Symp. No. 226}, in press.

\bibitem{Lin05}
Lin, R.P.:2005,`The Living With a Star (LWS) sentinels mission',
{\it In SPIE Proc.}, {\bf 5901}, in press.

\bibitem{Moses04}
Moses, J.D.: 2004, `A private communication'.

\bibitem{Newmark05}
Newmark, J: 2005, `A private communication'.

\bibitem{Tsurutani03} Tsurutani, B.T., Wu, S.T., Zhang, T.X., and Dryer, M.:
2003, `Coronal Mass Ejection (CME)-induced shock formation,
propagation and some temporally and spatially developing shock
parameters relevant to particle energization', {\it Astron. and
Astrophys.\/}, {\bf Vol. 412}, p.~293.

\bibitem{Tsurutani05} Tsurutani, B.T., Judge, D.L., Guarnieri, F.L.,
Gangopadhyay, P., Jones, A.R., Nuttall, J., Zambon, G.A., Didkovsky,
L., Mannucci, A.J., Iijima, B., Meier, R.R., Immel, T.J., Woods,
T.N., Prasad, S., Huba, J., Solomon, S.C., Straus, P., Viereck, R.:
2005, `The October 28, 2003 Extreme EUV Solar Flare and Resultant
Extreme Ionospheric Effects: Comparison to Other Halloween Events
and the Bastille Day Event', {\it GRL\/} {\bf 32}, L03S09.

\bibitem{Williams68}
Williams, D.J., Arens, J.R. and Lanzerotti, L.J.: 1968,
`Observations of Trapped Electrons at Low and High Altitudes', {\it
J. Geophys. Res.\/} {\bf 73}, p.~5673.

\bibitem{Ziegler03}
Ziegler, J.F., and Biersack, J.P.: 2003, `The Stopping and Range of
Ions in Matter (SRIM)', {\it www.SRIM.com}
\end{thebibliography}
\end{document}